\documentclass[12pt]{article}
\usepackage{latexsym}
\usepackage[english]{babel}
\usepackage{fancyhdr}
\usepackage[mathscr]{eucal}
\usepackage[dvips]{graphicx}
\usepackage{graphics}
\usepackage{color}
\usepackage{epsfig}
\usepackage{amsmath}
\usepackage{amsthm}
\usepackage{amsfonts}
\usepackage{amssymb}
\usepackage{amscd}
\usepackage{bbm}
\usepackage{latexsym}
\usepackage{marvosym} 
\usepackage{pifont}
\usepackage{subfigure}
\usepackage{psfrag}
\usepackage{caption}

\begin{document}

\title{ N-body problem with contact interactions}

\maketitle
G.F. Dell'Antonio

Math. Dept. University Sapienza (Roma)

and

Mathematics Area, Sissa (Trieste)

\vskip 1 cm \noindent
\section{Introduction }

We study  in the limit $ \epsilon \to 0 $  a system of $N\geq 3$ particles in $R^3$  which satisfy  the 
Schr\"odinger equation with  hamiltonian 

\begin{equation}
H^\epsilon (V)  = -\sum_{k= 1}^N   \frac { 1} {2 m_k} \Delta_k + \sum_{i,j = 1 \ldots N } V_{i,j}^\epsilon (|x_1 - x_j|)
\end{equation}

where  the potentials scale as $ V_{i,j}^\epsilon (|y|) = \epsilon^{-3} V_{i,j} ( \frac {|y|}{\epsilon}) $  and $ V_{i,j} (x) $ are  $L^1$ functions with  compact support (this last condition can be be substituted by the condition  of a sufficiently decay fast at infinity). 

The sum is in general over a subset of the indices. 

Notice that the $L^1(R^3) $ norm of the potentials does not depend on $ \epsilon$ and the potentials  
  converges \emph{weakly} (in distributional sense) when $ \epsilon \to 0$ to a delta distribution.  
Therefore in the limit $ \epsilon \to 0$ the "potentials"  are distributions supported by  some of the contact hyper-planes $ \Gamma_{i,j} \equiv \{ x_i = x_j \}$ . 

We take the forces to be attractive. 

The scaling of the potential we introduce \emph{is different} from the scaling  $V_N(x) = N^{3 \beta} V(N^\beta (x)$ which is used in studying the fluctuation  of $N$-particle quantum systems around the non linear Schr\"odinger equation (see e.g. e.g.[BCS] ) and \emph{also different}  from the scaling adopted in [A]to define \emph{point interactions}.

We will come back  to this difference.

We denote by $ \Gamma$ the union over the selected sets $ \Gamma_{i,j}$ and we will denote "hamiltonian for contact interactions" or simply \emph{contact interaction} the limit operator. 

 \emph{The limit  is described equivalently by  boundary conditions} at $\Gamma$ . 
 
 The equivalence can be seen by taking the scalar product with functions in the range of $H_0$ and integrating by parts.
 
 This step is crucial if one wants to see contact interactions as limits of interactions with smaller and smaller support. 

 This is the same procedure which provides different realizations e.g. of the laplacian on $ (0, \infty) $
 
 We will investigate resolvent convergence and spectral properties. . 

We consider also the case in which there are zero-energy resonances. 

With our scaling in the  limit  $ \epsilon \to 0$ the potential is formally a distribution supported by some of the \emph{coincidence} hyper-planes $ x_i = x_j .$ 

The limit hamiltonian is therefore a self-adjoint extension of  the free hamiltonian defined on functions that  have support away from the coincidence manyfold 

\begin{equation} 
\Gamma \equiv \cup'_{i,j} \Gamma_{i,j} \qquad \Gamma_{i,j} \equiv \{ x_i - x_j = 0 \} 
\end{equation} 

Functions in the domain have a singularity $ \frac {1}{|x_i - x_j|} $ at $ \Gamma_{i,j}$ for the given subset of indices (solutions of $H_0 \phi (x) = 0, x \notin \Gamma).$. 

Condition  of this type were assumed [B,P] by H.Bethe and R.Peirels in 1935 in their analysis of the proton-neutron interaction; later they were assumed for $N=3 $ by G. Skorniakov and K.Ter Martirosian  [S,T] in their study of the three body problem in Nuclear Physics in the  Faddaev formalism . 

For every $N$ we shall call them  Bethe-Peirels (B-P) boundary conditions .

Notice the condition $ N \geq 3$.  

Contact interactions describe the short range interactions between \emph{pairs} of particles (a particle may belong to two pairs). 

In the case $N=3$ the bound states, if present, are three-body bound states (trimers in the Theoretical Physics literature);  their wave function has support near the triple coincidence point. 

They may be infinite in number and then they are Efimov states.

For $N=4$ the negative point spectrum may have a contribution that correspond to a bound state of two pairs  (quadrimer);  their  wave function have wider  support. 

Also the quadrimers may be Efimov states.  

\bigskip
  Since in the limit the potential is a delta distribution, it is convenient to use an auxiliary  Hilbert spacein which the potential is more regular. 
  
  The auxiliary space is constructed with the aid of the positive operator $H_0 + \lambda$ where $ \lambda > 0 $ is an arbitrary positive number. 

Remark that contact interactions are \emph{a special class} of the extensions of the free hamiltonian $ \bar H_0$ restricted to functions with support away from $ \Gamma$ [P]. 

Since this operator is positive, the possible extension are classified by the theory of Birman, Krein, Visik [B][K][A,S].

This theory provides also information  on the regularity of functions in the  support of the extensions.

To analyze \emph{the specific extension we consider} (\emph{the only one that is limit of regular potentials}) we use a quadratic-form version of this method [A,S] and introduce an auxiliary space.

We call  it \emph{Krein space}  and \emph{Krein map} the (compact) embedding.

The (compact) embedding  into this space has the same role as the map from potential to charges in electrostatics and the auxiliary space is the counterpart  of the space of charges.

\bigskip	
\emph{Remark} 

The method does not apply to the case for $N=2$ the reason being that in the center of mass frame the contact set reduces to a point and there is no room for a compact embedding.  

The approximating hamiltonians have no limit. 

Still the equation of motion are well defined in the limit because the diverging term in the hamiltonian is a \emph{c} number .

The same problem is encountered in electrostatics: the potential due to a "point charge" is well defined,
but in order to define \emph{charge at a point}  one must consider the distribution of charges over a sphere of radius $ \epsilon$ and then let $ \epsilon \to 0$.

If the two-body hamiltonian has a zero energy resonance \emph{a different }  hamiltonian of "zero range interaction"  was introduced in [A] ; the resulting system is  called  \emph{point interaction}.

The limit hamiltonian is obtained  by scaling the potential as $ V_\epsilon (|x|)=  \epsilon^{-\frac {3}{2}} V(\frac {|x|} {\epsilon}) $ so that the $L^2$ norm does not depend on $ \epsilon $ and in the limit the $L^1$ norm vanishes. 

It is proved in [A] that this  Hamiltonian is self-adjoint. 

The difference in scaling is due to the fact that due to the  zero energy resonance  the difference of the resolvents (for the free and the interaction case) for the two-body problem  becomes singular as $ \epsilon^{-\frac {3}{2} } $ and this implies a difference in  scaling of the potential.

 The proof given in [A] requires the use of deep results   in Functional Analysis  because the missing factor in the scaling of the potential  (a \emph{short distance modification})   comes from the behavior of the wave function \emph{at large distances}. 
 
This requires a detailed analysis of the low energy singularities in the Birman-Schwinger formula for the difference of two resolvents. 
 
 Notice that for $N \geq 3 $ the presence of a zero energy resonance in the two-body potential does not lead to singularities in the Birman-Schwinget formula and the Krein map is well defined. 
 
 We shall comment in the Appendix on a possible relation between contact and point interactions.

  \bigskip
  
  .........................

\section{ Analysis in the auxiliary space}

The  \emph{Krein map}  (compact embedding) is achieved acting with  $ (H_0 + \lambda)^{- \frac {1}{2}}$. 

The Krein space could be called, in analogy with electrostatics, \emph{space of charges} and the Krein map can be compared with the  map from potentials to boundary charges in electrostatics (notice that the operator that induces the map is a first order differential operator). 

The embedding has the advantage that the boundary potentials are now less singular (they are $ L^1$ functions) and convergence when $ \epsilon \to 0$ is  easier to prove. 

\bigskip
If zero-energy resonances are present  the condition $\lambda > 0$ is required for compact embedding . At the end we shall undo the Krein map, so that the actual value of the positive parameter $ \lambda $ is irrelevant. 

For the sake of simplicity we shall occasionally set $ \lambda = 0 $ in our formulae so that the the kernels of the two forms become homogeneous of order $ -1$ in the coordinates. 

This helps greatly in keeping the formulae simpler and does not alter the short distance behavior.  

In Krein space the potentials are $L^1 $ functions and the presence of a zero energy resonance does not alter the domain. 

 We can consider therefore  in Krein space also the case in which there are  zero energy resonance obtained e.g. by monitoring  magnetic fields (Fesbach resonances) .

Recall that a zero energy resonance in a two-body system with a potential invariant under rotation is defined, in relative coordinates,  as a rotational invariant solution $ \phi (x) $  of $ \Delta \phi (x) = 0 , \; x \not= 0$ .

One has therefore  $ \phi (x) = \frac {a}{|x|} + b , \;\; a, b \in R $.

We require $ b=0$ ; in this case $4 \pi a $ represents  the \emph{scattering length}   which is  by definition 

\begin{equation} 
a  \equiv  lim_{R \to \infty} \frac {1}{R} \int _{ |x| < R} |\phi (x) |^2 d^3 x 
\end{equation}

\bigskip
One has then 

\begin{equation} 
\phi(x) =  a \int_{ S^3} d \hat k  \int _0^\infty d \rho   \frac {e^ {i \hat k. \rho \hat x} }  {\rho} 
\end{equation} 
 
\bigskip
We denote $K_M$ the operator \emph{in Krein space} which is image of the hamiltonian.  

The operator $K_M$   has been used  by Minlos $[M_1][M_2]$ in his analysis of contact interactions for $ N=3$.  

In Krein space  the kinetic energy is represented by the operator $ \sqrt { H_0 + \lambda} $ and the (negative) interaction potential is an $ L^1(R^3)  $  function which is the limit of functions that represent tin Krein space the potentials $ V_\epsilon$.  

This makes rather easy in Krein space the proof of resolvent convergence when  $ \epsilon \to 0$. 

In this space one has a Birman-Schwinger formula  for the difference of resolvents.

The image of the Hamiltonian in Krein space  the difference of two unbounded positive self-adjoint operators associated respectively to the kinetic energy and  to the potential.

Depending on the masses and the strength of the potential the resulting operator may be self-adjoint ( "regular" case) or only symmetric (singular case) .

In the singular one has \emph{in Krein space}  a family of self-adjoint operators (\emph{not a direct sum}) 

We will see that this  is an instance of the Weyl limit-circle case. 

One has now to "come back"  to  "physical space". 

In the regular case  the proof of convergence in physical space is done by using the compactness of the map and  the fact that weakly closed  \emph{positive} quadratic forms are strongly closed.

In the singular case we decompose first  the quadratic form  using the symmetry under rotations and then make use of $ \Gamma-$ convergence [Dal]  on the resulting forms (that are locally strictly convex). 

In both cases in each sector one ends up with a \emph{unique self-adjoint operator} in each channel.

In the singular case the resulting operator \emph{in physical space} has a negative point spectrum;  it may  consists of infinitely many points,  accumulating geometrically at zero. 

This "Efimov effect" \emph{is not related} the the effect with  the same name  which occurs for regular potentials with  zero-energy resonances.

In the present case the effect is \emph{independent of the presence of zero energy resonances}. 

The spectrum of the N-body system with contact interactions depends on the masses and on the statistics.  

We will prove that  the (negative) point spectrum \emph{is completely determined by the three- and four-particles sub-sectors}. 

We stress that both in the regular and in the singular case the self-adjoint operator that represents the contact interaction is the limit, in strong resolvent sense, as $ \epsilon \to 0$, of the sequence of operators $ H_\epsilon = H_0 + V^\epsilon$. 

\bigskip

We describe at length the cases $N=3$ and $N=4$ and more briefly the generic $N$ case. 

We analyze in detail two cases: 

1) a pair of identical scalar fermions of unit mass interacting with a third particle of unit mass.

2) a pair of identical spin $ \frac {1}{2}$ fermions in contact interaction . 

In both  cases the hamiltonian is positive. 

In the second case contact interaction takes place between electrons with opposite spins (Cooper pairs). 

There is an easy generalization to the case of $N$ pairs of spin $\frac {1}{2} $ fermions  (unitary gas)

\section{Boundary conditions}

In the classical case  \emph{constrained systems} are defined by restricting the configuration      .  

This is often attributed to the action of very strong constraining forces (perpendicular to the constraint manyfold  if the constraints  are \emph{perfect}). 

In Quantum Mechanics contact interactions can be  defined   by imposing boundary conditions at  the coincidence manyfold $ \Gamma $ i.e. selecting a self-adjoint extension of the restriction of $H_F$  to functions supported away from the  \emph{coincidence manyfold } $ \Gamma$ .

\begin{equation} 
\Gamma = \cup_{i,j} \Gamma_{i,j} ,\;\; \Gamma_{i,j} \equiv \{x_i-x_j =0 \} .
\end{equation} 
 
where the union is taken over indices that correspond to constraints. 

We work in the reference frame in which the barycenter is at rest  and therefore in $ L^2 (R^{3(N-1)}) $ (the \emph{physical space}) . 

\bigskip
The  boundary  conditions  correspond  \emph{formally} to potentials $ v_{i,j}$ that are  \emph{distributions  supported by the boundary}. 

They are expressed by the requirement that at the boundary functions in the domain have at most the singularity

\begin{equation} 
\phi (X) =\sum_{i \not j}  \frac {C_{i,j} } { |x_i-x_j|} + b_{i,j} + 0 (|x_i - x_|) \quad i \not= j  
\end{equation}

 where  the constants  $ C_{i,j} $ characterize the boundary conditions. 

\bigskip
We shall prove that the hamiltonian for contact interactions is  limit in strong resolvent sense of hamiltonians with regular  two-body potentials 

The advantage in using contact interactions is of a simpler mathematical formulation and independence of the detailed form of the approximation  potentials. .

\bigskip
We shall prove that if no further interactions are introduced, for contact interactions the negative part of the spectrum of the hamiltonian of the $N$ body system is \emph{completely determined} by the spectral properties of the  three- and four-body subsystems. 

Notice that  for smooth potentials it is well known  that  \emph{pairs of zero energy  resonances contribute additively to the resolvent of the operator} ( a consequence of the Birman-Schwinger  formula for the difference of two resolvents) .

This property holds also for contact interactions. 

\bigskip
\emph{Remark 1}

The difference between resonant  and non resonant contact interaction is seen when there are bound states in the three and  four body channels. 

Since the Krein map is order preserving it is sufficient to consider the order in Krein space, where the contribution of the zero energy resonances are not masked by singularities of the potentials. 

The presence of a zero energy resonance increases the contribution of the potential part.

Therefore one expect that the bound states have lower (negative) energy.

Small discrepancies between  predicted and observed values have been reported [C,M,P] [Ba, P]. 

The predicted values are obtained for $ K_M$ (the resolvent has a simpler form) and this may be the cause of disagreement. 

The difference of the spectra of contact interactions between the resonant and non-resonant case is small if the bound states are sharply localized, but it may become relevant if their essential support is large (this occurs e.g. in the Efimov effect). 

\bigskip
.....................................
 
\bigskip
\emph{Remark 2} 

 The Krein map  depends on $N$ and on the masses of the particles (the "potential"  is essentially a convolution of the potential  at the boundary in physical space with a Green' s function) .
 
 Therefore \emph{in Krein space} the boundary conditions  depend  on the position, masses and  symmetries  of all the particles. 

They are given now as 

\begin{equation}
\phi (X) = \frac {a_{i,j} (Y )   }{ |x_i-x_j|} + b_{i,j} (Y) + C(Y)  f(|x_i-x_j|) \quad i \not= j  \quad f(0) = 0 
\end{equation}

where $Y$ represents the other differences of variables and $A \equiv \{ a_{i,j} \}$ , $ B \equiv \{ b_{i,j} \} $ 
are suitable functions at the boundary.  

The functions $ a_{i,j} (Y) $ and $ b_{i,j}(Y) $ 
depend on the masses of all the particles.

There are choices of the masses for which the matrix $ A \equiv a_{i,j}(Y) $ is singular (in some cases as singular as  $ \frac { 1}{ |y_i | log |y_i | } $ in each  variable).  

Notice that these singular  boundary conditions occur in Krein space;  \emph{in physical space the boundary conditions are the usual ones}. 

\bigskip
  We  will call \emph{homogenous} the boundary  conditions if the matrix $ B \equiv \{b_{i,i} \} $ is absent.

The case $  B \not= 0$ corresponds, in both the classical and the quantum settings, to the presence of magnetic forces at the boundary. 

\emph{In what follows we consider exclusively the homogeneous case}. 

 It corresponds in the classical setting  to requiring that the constraints are \emph{perfect } (the constraining forces are perpendicular the to the manyfold).

\bigskip
.........................................

\bigskip
\emph{Remark 3}

A stronger modification of the hamiltonian  occurs if some of the  elements $ a_{i,j}$ has a singularity $ \frac {1} {|x_i - y_k|}  $ so that the function at the boundary has singularity $\frac  {1}{| x_i - x_j ||x_i - x_k| } $ for  different values of the  indices.

 In [M,F] the authors consider this case for $N=3$ . It describes \emph{in physical space} three particles in simultaneous contact interaction.

Also in this case  there is a family of dynamics (self-adjoint extensions). 

Each self-adjoint extension  is unbounded below \emph{in  physical space}  and  has infinitely many eigenvalues $ \mu_i ,  i \in Z.$  which diverge geometrically to $ -\infty$. 

\bigskip
These extensions \emph{cannot be recovered}  as limit of hamiltonians  with smooth potential $V^\epsilon_{i,j}  (|x_i - x_j| ) $ which belong uniformly to $L^1$, have  support of volume $ \epsilon ^3 $ and are  scaled so that the  $L^1$ norm stays constant.   

\bigskip
.............................

\bigskip

\section{The cases N=3 and N=4 }. 

\bigskip
$N=3 $

For three particles  we will consider only the case in which a  particle is in contact interaction with the other two while these two particles do not interact .

For the sake of simplicity we consider the case in which there are enough symmetries, e.g. the two non interacting particles are either a pair of identical fermions or a pair of identical bosons.

In this case the  the quadratic form has  kernel in $R^3$. .

We restrict first our attention to the case in which two of the particles are identical fermions of mass 1 and the different particle has mass $m$. 

We shall later treat briefly the case of two identical bosons.

We shall prove that in Krein space the quadratic form is the sum of a  positive quadratic form $K_0$ (image of the kinetic energy) and of  a form that  in configuration space has kernel $K_1$.

The kernel $K_1$  is the sum of  $  -\frac {C_j(m) }{ |x_j- x_0|} $  with $C_j(m) > 0$ ( $x_j$ are the coordinates of the fermions  and   $x_0$ is the coordinate of the third particle)  and of a \emph{positive}  bounded kernel which vanishes for $ x_i = x_0$.

The constant $ C_j(m) $  depends on the mass of the particle and on the symmetries of the system. 

There is a constant $\bar C  > 0$  such that if $ C_j \geq  C  $ the operator associated to the form $ K_0    -\frac {C_j(m) }{ |x_j- x_0|} $  is symmetric but not self-adjoint.

It is in the Weyl limit-circle case and  has a one-paramter family of self-adjoint extensions. 

Each member of the family has negative pure point spectrum; for some values of the parameters the spectrum may extend to $- \infty$ geometrically scaled.

There is a member of the family that has the lowest spectrum; the corresponding  eigenfunctions have a $ \frac { 1}{ |x|^{\alpha(m)} } $ behavior at the singular point ($\alpha $ is such that the function is in Krein space) .

The same holds true for the operator  associated to the form $ K_0 + K_1$. 

Recall that this statements \emph{hold true in  Krein space}

\bigskip
$N=4$

For four particles the "kinetic part" depends in Krein space on all four masses.  

There are two types of contributions to the potential part corresponding to the following configurations

\bigskip
a)
one particle does not interact, and we are back to the case of three  particles.

\bigskip 
b)
 there is an interaction between  two pairs particles  
 
\bigskip
For generic values of the masses and generic symmetry the quadratic form may have a complicated structure. 

In the case of two identical spin $ \frac {1}{2} $ particles we can take advantage of the symmetries; will prove that the kernel of the "potential part" of the quadratic form in Krein space when written in configuration space is the sum of a form that represents the operator of $  - \frac {C_2 }{ |y_1 - y_2|}$  and a locally bounded positive quadratic form and with kernel that vanishes for $ y_1= y_2 $. 

Here $y_k , \; k=1,2$ are the  coordinates of two particles of opposite spin and the coefficient  $C_2 > 0$ depends on the masses. 

Notice that this singularity in the quadratic form \emph{is not the image in Krein space of delta potential at coincidence points}.

It is rather a secondary effect due to the contact interaction between pairs. 

Consequently the four-body bound states (in Krein space and then in physical space) have eigenfuctions with large support  and correspond to  \emph{interaction between barycenters}.

Also in this case if the coefficient $C_2$ is large enough we are in the Weyl limit case the form corresponds to a family of self-adjoint operators each with negative spectrum .

For each angular momentum sector  there is a distinguished extension which has the lowest spectrum. 

If  $C_2$ is still lager the (negative) pure point may extend geometrically to $ - \infty$ ; it produces  \emph{in physical space} an Efimov effect. .

\bigskip
This leads to a natural analogy.

Recall that in  Krein space each pair of particles in contact is represented by a "charge distribution" . 

The singular potential  term can be regarded as a Coulomb potential  which represents the interaction between the two charge distributions. 
 
It has therefore the same structure of the term described in the case of three particles \emph{but the singularity is   on a different manifold in configuration space}.

The final result for the four body case in Krein space is the sum of the "kinetic energy" , a positive form locally bounded and  two (negative)  contributions which have a Coulomb-like singularity in different points in configuration space.

\bigskip
As in the three particle case whether in Krein space there is a unique self-adjoint extension or families of self-adjoint extension depends on the relative weight  of the positive(kinetic energy)  and negative (boundary potentials) parts of the quadratic form.

\bigskip
We are therefore , both in case of three particles and in the case of four particles, reduced in Krein space  to the analysis of a symmetric operator that differs from  the sum of $ \sqrt{ H_0 + \lambda} $  with  a negative "Coulomb potential" by a bounded positive quadratic form with kernel that vanishes on the diagonal. 

This sum has been studied in much detail [lY][D,R] in the context of the relativistic hydrogen atom. 

We shall make use of their findings. 

The problem can be analyzed separately in the angular momentum sectors and the limit cycle property is not affected by the addition of the positive regular  part.  

Using the results of [lY] and especially of [D,R] we shall prove that, depending on the value of the masses and the symmetries, the quadratic form we are studying,  both  for three and four particles, either is the quadratic form of a self-addiont operator or it represents (\emph{in Krein space}) the quadratic forms of a family of self-adjoint operators with negative non degenerate point spectra.

For some value of the masses the negative part of the spectrum  (in Krein space) extends geometrically  to $ - \infty $.

The resulting form has the same structure for $N=3 $ and $N=4$ but the (Coulomb) singularities are in different variables and produce separate Weyl limit-circle effects. 

We stress again that the Coulomb singularities  between pairs of particles in contact are found in Krein space \emph{are not image of further contact interactions} and the wave functions  in physical space {are not supported near the near a quadruple coincidence point}.

\section{Results in physical space}

The analysis we have done so far is of the system in Krein space.

In order to have the description in "physical space" one must invert the Krein map. 

In Krein space the forms associated to $\tilde H_0 + V_\epsilon $ converge strongly when $\epsilon \to 0$ to the form associated to the contact  interaction.

Compactness of the Krein map implies weak convergence in physical space.

If the limit form is positive and closed  (the regular case)   weak convergence implies strong convergence in physical space on the domain of the limit operator and then strong resolvent convergence. 

The domain of the resulting self-adjoint operator is obtained (by duality) acting with the Krein map on the domain of the operator in Krein space. 

The structure of the Krein map implies that in this "regular case" the functions in the domain of the limit operator in physical space satisfy the B-P boundary conditions. 

 In the \emph{singular case} the limit form is not closed. It corresponds to a family of self-adjoint operators \emph{ but not to a direct sum}.

In this case the convergence of the hamiltonians in physical space is obtained by $ \Gamma-$convergence [Dal].  

Due to rotation invariance one can consider separately the sectors of fixed angular momentum.

By duality the inversion is obtained by applying the Krein map  to the domain of the family of self-adjoint extensions. 

It is easy to verify that in each angular momentum sector the quadratic form in physical space is locally strictly convex and satisfies   the conditions for $ \Gamma-$convergence (essentially lower semicontinuity).  

Recall that the $ \Gamma$-limit is the unique form $F$  such that for any sequence the following holds 

\begin{equation}
\forall y\in Y ,\; y_n \to y  \;\;  F(y) \leq lim  inf_{n \to \infty} F_n (y_n)  \;\;\; \; \; lim sup_{n \to \infty} F_n (y_n) \geq F(y) 
\end{equation} 

Therefore  in the singular case, for $N= 3$ and $N=4$ and any choice of the masses  there is in physical space \emph{a distinguished extension} (the $ \Gamma-$limit) obtained by $ \Gamma$-convergence.

Since the Krein map preserves order, the $ \Gamma-$limit  coincides with the lift to physical space of the self-adjoint extension (in Krein space)  that has eigenfunctions which have  power-law singularity. 

One can at this point set to zero the arbitrary parameter $ \lambda $ since in the limit $\lambda \to 0$ there are no infrared divergences (due to convolution with the Green's function). This extension  has  bound states;  their number depends on the masses.

\bigskip
For quadratic forms which are bounded below $\Gamma$-convergence implies strong  resolvent of the associated operators [Dal].

Therefore 

\bigskip

\emph{ Proposition}

\bigskip
{Contact interactions are limits in the strong resolvent  sense, of hamiltonians with potentials of shrinking support }.

\bigskip

.............................

\bigskip
This implies, among other things, that bound states converge to bound states, and in particular that contact interaction is a good tool to find (approximately) the location of the bound states for potentials sharply peaked at the origin.

In the case when there is an infinity of bound states their energies geometrically scaled and  their support is increasing, as in Efimov effect.

Notice that this Efimov effect for trimers and quadrimers \emph{is not related}  to the Efimov effect due, in case of regular potentials, to the presence of zero-energy resonances in at least two channels in a three body problem.

In fact, in the present  case, this Efimov effect \emph{is independent of the presence of two-body resonances}.

The presence of resonances does not alter the general picture but leads to a difference in the position of the Efimov bound states .

\bigskip

\emph{Remark 1} 

The "tail" of the Efimov states (small energy states) are hardly present in a realistic model, which keeps into account other type of interactions and corresponds to $ \epsilon $ very small but not zero.

What can be seen in experiments are a few  members of the "head" (low energy  states)  of the Efimov sequence; they  should be recognized as Efimov states  for the geometrical scaling property of the energies. 

This states are less affected by the other short range interactions.

On   this states the effect of the zero-energy resonances might  be  more visible.

Three  body Efimov states  have  probably be seen experimentally [Pe], [C,M,P] and also four-body Efimov states have been reported [C,M,P][Ba,P].

\bigskip
.......................

\bigskip
\emph{Remark 2} 

The approach of  $[M_1][ M_2] $(see also $[C_1],  [C_2]$) for the case of three particle follows essentially the lines described here . 

The choice of the auxiliary space  (and therefore of the Krein map) is not stated explicitly  but can be inferred  by the fact the image of $ H_M + \lambda$  is its square root. 

The analysis is done  in Fourier space and the behavior at the coincidence   hyper-planes (which is  easy to describe in configuration space)  is imposed as  specific bounds "at infinity" on an integral in Fourier space. 

For the case in which the form is not positive a one-paramter family of self-adjoint extensions  in Krein space is found but since the analysis is done in momentum space  the relation  with Weyl's limit circle case is not  recognized.

In $[M_1]$ the multiplicity of extensions  is related to the multiplicity  of solutions of the algebraic equation that expresses the condition under which an operator is self-adjoint . 

The inverse power law behavior at triple coincidence  points of the wave functions of the bound states in Krein space is correctly stated  in $ [M_1]$but is origin is not discussed.

The need of "coming back" to physical space and  the presence of Efimov states is correctly stated in $[M_1][ M_2] $ but the uniqueness of the limit is not discussed . 

Also resolvent convergence is not stated.

.\bigskip
.........................

\bigskip
\emph{Remark 3}

We have so far considered only the case in which the matrix  $B$ with matrix elements $b_{i,j}$ in the boundary conditions at $ \Gamma$  is set to zero. 

This term is originated by a magnetic potential (dependent linearly on the momentum) it may lead to other bound states.  

We don't consider here this interesting problem  

\bigskip
.............................

\section{N-body sistems}

We have considered so far the cases $N=3 $ and $ N=4$.

We will see that for the general $N$-body problem the negative part of the spectrum is pure point  and its structure can be reduced to these two cases, $N=3$ and $N=4$. 

We shall indeed prove that in the $N$ body case the negative part of the spectrum of each one of the extensions  is contained \emph{in the algebraic  union}  of the point spectra of its three- and four-particle subsystems. 

This may be seen as a consequence of the fact that in Krein space the "kinetic part" is a first order differential operator and it has zero newtonian capacity.

Since this holds for any four-body subsystem its contribution is independent of the presence of any other particle. 

In particular a system composed of $N$ spin $ \frac {1}{2} $ fermions is stable (its spectrum is the positive real axis).

If something is known about the commutator  of the operators associated to these subsystems, more can be said about the spectrum.

Indeed \emph{in physical space} the potentials are supported by the coincidence hyper-planes and a four-fold coincidence sets gives no contribution.  

In a four particle system there is a new contribution  that can be seen  in Krein space  as due to  a first order differential operator describing \emph{ a closed system } ( two pairs of particles  interacting through a potential field). 

 In Krein space the potential on each pair  has a tail $ \frac {1}{|x_1 - x_2|} $ and the overlap of two tails  is sufficient to bind the two pairs.

This  effect is not altered by the presence of a fifth particle. 

Since these are the only potential which act on an isolated four-body system one has 

\bigskip
\emph{Proposition }

 \emph{For contact interactions of $N$ bodies the \emph{negative}  point spectrum (both in Kein space and on physical space) is completely determined by the three- and four -body subsystems}

 \bigskip
 ........................

\bigskip
Graphically one can describe  this by saying that the contact interaction among $N$ particles can be visualized by a diagram in which there is a link between two points if there is contact interaction between them. 

The diagram is made of $V$ shaped components, corresponding to  the configurations in which one body is  in contact interaction with two other bodies, and   an $H$ shaped components representing Coulomb interaction between two pairs each in mutual contact interaction.

\bigskip
Something more can be said if one has information on the commutation properties of the operators associated to the components.

For example in the case  of contact interaction of a system composed of a particle of mass $m $ and $ N$ identical scalar fermions of mass one,  since there is no contact interaction between the fermions,  the system decouples in subsystems of three particle with no  interaction with the remaining particles. [M,S] 

Another favorable case is the one in which all subsystems are stable.

This is the case  for a system of $N$ pairs of spin $ \frac {1}{2} $ identical fermions (unitary gas);  in fact we have proved that in this case all three and four particles subsystems are stable. 

In the previous section we indicated  that  the negative part is contained in the union the negative spectra of its three- and four particle compnents. 

The quadratic form is positive for a system of two pairs of spin $ \frac {1}{2}$.

It follows that the quadratic form of a system of an arbitrary number of pairs of $ \frac {1}{2} $ equal mass fermions is positive.  

In the Theoretical physics literature this system is called \emph{unitary gas}. 

Therefore that the unitary gas is stable. 

\bigskip
\emph{Remark  1}

From the analysis above it follows that there  can be  a three or four particle Efimov effect but there is no five body Efimov effect. 

If there are other interactions this last statement must be taken with care.

For example one may have an Efimov effect [Pe] for a system of  four fermions and a light particle.

This can be seen as follows: two pairs of fermions are bound  \emph{due to other short range forces}  and the two \emph{resulting bosons} interact with the other particle through a contact interaction. 

  ...........
   \bigskip

For an outlook on experimental and theoretical results on the three and four body problem one can consult [C,M,P] [C,T] [Pe].

\bigskip
\emph{Remark 2} 

For regular potentials pairs of resonances contribute \emph{additively} to the resolvent.

Due to strong convergence this is true also for resonant contact interactions.

Therefore also in the resonant case the negative part of the spectrum  of the self-adjoint operators describing the system \emph{is completely determined by the two- three and four-body subsystems.}  

   \bigskip
   .............................

\section{Other systems} 

The Krein map can be used in other problems. 

A model example is the one dimensional Salpeter (semi-relativistic Schr\"odinger) equation [S,B]  with an interaction localized at the origin \emph{i.e. defined by boundary conditions at the origin}. .

For $N=2$ the model  is analyzed  in [A,K] as "point perturbation" of a pseudo-differential operator. It is discussed in some detail in [A,F,R] and in [K,R] from a more formal point of view.

An attempt to see this interaction as due to a \emph{renormalized }$ \delta$ has been done in [E,T]. 

In [A,K] it is proved that the symmetric operator $ \sqrt{-\Delta  +1 }$ , defined on smooth functions with compact support away from the origin,  has deficiency indices $ (1,1)$ and therefore admits a one parameter family of self-adjoimt extensions. 

The Authors give explicitly the deficiency subspaces (and also the scattering structure).

There is  one family of self-djoint extensions; the "boundary conditions" at the origin are now 

\begin{equation} 
\psi (x) = c_\psi log |x| + \psi (0) + o(1) \qquad  c_\psi \in C 
\end{equation}

For generic values of $N$ one  can perform the analysis of the N-body problem of the Salpeter model  with contact interaction following the scheme we used for the Schr\"odinger case.

The boundary behavior is more regular but now the differential operator is of first order.

The Krein map is given now by the operator $ (\bar H   + \lambda )^{-\frac {1}{4} }  $ 

It has as usual the effect of smoothing the distributional "potential" at coincidence hyperplanes but at the same time it lowers the differential structure of the positive part which is now a (pseudo)-differential operator of order $ \frac{1}{2}$.

In Krein space the quadratic form is now 

\begin{equation}
(\phi,  ( H_0 + \lambda )^{\frac {1}{4} } \phi)  - (\phi, \Xi _\lambda \phi)  
\end{equation} 

where the operator $ \Xi $   the convolution of  $ (H_0 + \lambda )^{-\frac {1}{2} } $  with the delta distributions at the coincidence hyper-planes.  

Depending on the choice of the masses this quadratic form is positive or it has logarithmic singularites.

In the latter case the form is not closed and the symmetric operator can be decomposed in a family of self-adjoint operators with simple negative point spectrum.   

 Also here on goes back to physical space undoing the Krein map. 

This is done, as in three dimensions,  by  compactness on the positive  part of the spectrum and by $ \Gamma-$convergence for the  singular part. 

Since the form is bounded below, the Efimov effect does not occur

\section{Details for the three body case} 

We shall first treat in detail  the case of two spinless fermions of mass one in contact interaction  with a third particle of mass $m.$ 

We shall then briefly consider the case in which the two identical particles are bosons.

\bigskip
In this section we survey and generalize known results [Pa] $ [M_1] [M_2 ]$ see also $ [C_1] [C_2]$.

This references use $ H_M  $  as a reference operator (i.e. we are in the no  zero energy resonance case) .  

In the  next section  we  provide a simple model \emph{in Krein space } that has the same multiplicity of self-adjoint extensions with the same mass thresholds 

This model has been studied in the  past [lJ][D]; it is related to the relativistic Coulomb problem.  

Remark that, due to the antisymmetry of the wave function under permutation of the coordinates of the fermions,  one can consider only  a single   boundary potential.

\bigskip
 In $[M_1][M_2]$ the Author follows ,at least implicitly,  the approach to self-adjoint extensions that we have called Krein map.

 It is proved   that  the  system  is described \emph{in Krein space}  by a symmetric  operator $\hat Q$ sum  of two operators $\hat Q_i , \; i=1,2 $ .  

\begin{equation}
\hat Q= \hat Q_1+  \hat Q_2
\end{equation} 
 
The quadratic forms $Q_1 $ and $Q_2$ of the symmetric operators are given (explicitly in [Pa] and implicitly in $ M_1, M_2 $).

Taking into account the antisymmetry requirement in $ [M_1]$ one is reduced to study in Krein space the forms (for $ \lambda = 0$)
 
 \begin{equation}
 Q_1(\phi)  =  \frac {m}{m+1} (\phi, \sqrt {-\Delta } \phi) 
\end{equation}

and $Q_2$ with kernel 

\begin{equation}
Q_2(p,q) = \frac { \frac{2}{1+m} (p.q) } { (p^2 + q^2)^2 - \frac {4}{ (1 + m)^2 } (p.q)^2 } 
\end{equation}

These are precisely the quadratic forms  \emph{in Krein space} that correspond respectively the kinetic energy and the contact interaction.

\bigskip
While the forms  $  Q_1$ and $ Q_2 $ define uniquely self-adjoint operators $  \hat Q_1$ and $ \hat Q_2 $, the operator $ \hat Q $ associated to  $  Q $ is a priori only symmetric (it is the \emph{form sum} ). 

One verifies that the operator $ \hat Q_1, \; \hat Q_2 $   are invariant under rotations and therefore can be analyzed separately in each angular momentum sector (and therefore the same is true for their sum). 

The mass $m$ of the third particle is the only parameter . 

\bigskip
\emph{Recall that this analysis is done in Krein space}.

\bigskip
Denote by $Q_l $ the sum of the projections  to the sector of angular momentum $l$

This operator can be diagonalized by a Mellin transform.  

In $ [M_1][M_2] $  the Author proves that there are  constants $ m^{**}_l $ and $ m_l^* $  such that $Q_l$ s positive for $ m>m_l^{**}$ and therefore there is a self-adjoint operator ( the \emph{form sum} ) associated with it. 

If $ m \leq m^{**}_l$ the form $Q_l$ ceases to be positive .

While the two parts correspond to a self-adjoint  operator their sum is the quadratic form of an operator that is  symmetric but not self-adjoint .

The quadratic form is no longer closed. 

It is at this point that one sees the advantage of working with quadratic forms.

We shall prove that  the symmetric operator associated to the form can be  "disintegrated" into a family of self-adjoint operators with negative point spectrum. 

The the bound state(s) have eigenfunctions  that differ only by scale.

These \emph{three-body bound states} are supported near the origin (in the center of mass system). 

\bigskip
For each extension the space of bound states  is one-dimensional for  $ m^*_ l  < m \leq m^{**} _l $ and infinite dimensional for  $ 0 < m \leq m^*_l $. 

In this case the energy of the bound states accumulates (geometrically)  at minus infinity (Thomas effect).

Recall that the statement about the Thomas effect \emph{is made in  Krein space}. 

To come back to the "physical space" $ L^2 (R^{6)} )$ one must undo the Krein map. 

This is not stressed in $[M_1], [M_2]$. In particular in the singular case the use of weak sequential convergence is not mentioned

The map back to physical space changes the topology  and now the  eigenvalues of the three-body bound states accumulate at zero and   the support of the eigenvectors increases  as the eigenvalues converges  to 0 .

\bigskip
The presence of infinitely many extensions was noticed first by Danilov [Da] (see also [Pa]). 

In these references there is no explicit mention that these extension are defined  is a space of singular functions (in our case the Krein map  is a change in metric from $L^2 $ to ${ \cal H}^{- \frac{1}{2}}$).

It is  demonstrated  in $ [M_1][M_2]$  by finding in this space infinitely many solutions of   the algebraic equation $ H= H^* .$  

In the next section we will show how to obtain this result  comparing the energy operator \emph{on Krein space} with  the hamiltonian of the relativistic Coulomb model $ \sqrt{-\Delta} - \frac {C}{ |x| } $.

 This  hamiltonian has been thoroughly investigated [D,R], [lY] .
 
 In this model the plurality of extensions for $ m \leq m^{**}_l  $  is clearly seen  as a \emph{ Weyl limit-circle effect}. 

\section{Estimates in configuration space} 

Since one must compare the form associated  to the potential with $ \sqrt {H_0}$ and since for $ N \geq 3$ the diagonal form is not available,  it is convenient to write the "potential term"  in a form convenient to an analysis in configuration space. 

This will help also in a better understanding of the nature of the spectrum.

One has then 
\begin{equation} 
-\frac {1+m } { m } \frac {1} {(p-q)^2 } + \Xi (p,q,\lambda) 
\end{equation}
where $ \Xi $ is a positive kernel with  $ \Xi (p,p,0 ) = 0$

Therefore in Krein space the  operator is
\begin{equation}
2 \pi^2  \frac {m}{m+1} \sqrt {- \Delta} - \frac {4\pi(1+m)}{m} \frac {1}{ |x| } + \tilde  \Xi 
\end{equation} 

where $ \tilde \Xi $ is  a positive operator  with  smooth kernel. 

In the case of the fermions one must first anti-symmetrize the potential form and then write the kernel as function of the variable $ p-q  $ and $p+q$.

The resulting symmetric operator is 

\begin{equation}
2 \pi^2  \frac {m}{m+1} \sqrt {- \Delta} - \frac {1+m}{8 \pi(2m + m^2)} \frac {1}{ |x| } + \tilde  \Xi '
\end {equation} 
 
where $ \Xi' $ is a positive operator with locally bonded kernel vanishing on the diagonal.

Notice the the coefficient of the second term diverges as $ m \to 0$ both for fermions and for bosons. 

It follows that the form in Krein space should be compared with the quadratic form of the symmetric operator   

\begin{equation} 
\sqrt {- \Delta} - C(m) \frac {1}{ |x|} 
\end{equation} 

where $ C(m) $ is a positive function of the parameter $m$,

This function is different  in the bosonic and in the fermionic case but in both cases increases  monotonically to $ + \infty$ as $ m \to 0$.

Symmetric operators of the form  $ \sqrt { - \Delta } -\frac {C(m) }{ |x|} $ have  been studied extensively ([lJ][B,R] ) as a function of $C(m) $, originally  in the  context of the non relativistic hydrogen atom.

We denote them by $H_R $ (relativistic  Coulomb atom).

For these operators there are threshold values $ C^*, \; C^{**} $ such that for $ C > C^{**} $  the spectrum is absolutely continuous and positive. 

For $ C^* < C \leq C^{**} $ there is a continuous family of self-adjoint extensions, each with a negative  eigenvalue, and for $ = < C \leq C^* $ the negative spectrum is pure point and accumulates geometrically  to $ - \infty$ (a Weyl limit circle effect) .

In the latter case the eigenfunctions concentrate at the origin (eigenvalues and eigenfunctions are known explicitly) 

These results can be transcribed in function of the mass $m$ of the third particle and lead to different mass intervals in the bosonic case and in the fermionic one. 

In particular denoting by $ m^*_l,\; m^{**}_l$ the mass thresholds in the fermionic case one has $ 1> m^{**}_l$  for every $l$ whereas in the bosonic case one has $ 1 < m^{*} _0 $.

Of course $ M_l* = m_l^*, \quad  M_l^{**} = m_l ^{**} $.  

Since the eigenfunctions are known it is possible to show that the operators $\tilde  \Xi '$ and $\tilde \Xi $ are small on the eigenvectors  of the self-adjoint operators associated to $ \sqrt { - \Delta } -\frac {C(m) }{ |x|} $.

In the case $ m < M^{**}$ one can use rearrangement inequalities and   regular perturbation theory with convergent series to prove that  also for the complete forms the same  multiplicity of self-adjoint extensions occurs with the same thresholds holds. 

This approach has the merit to point out the role of the limit circle property and does not rely on an explicit diagonalization, which  is not known for $ N > 3$. 
 
\section{Two identical boson and a third particle} 

We now consider briefly the case of two identical bosons of mass 1 and a third particle of mass m.

In the case of two identical boson in Krein space the part of the energy form that comes form the boundary conditions is

\begin{equation}
Q_2(p,q) '= \frac {- \frac{ 2}{1+m} (p.q) } { (p^2 + q^2)^2 - c_f(m)  (p.q)^2 } \qquad c_f(m) = \frac {4}{ (1 + m)^2 }
\end{equation} 
 
 Now  the quadratic form is the sum  of a positive form $ \Xi''$ and of   $ \sqrt { \Delta } - \frac {C_b(m) }{|x|}$ where $ C_b (m) > C_f(m)$ ($b$ for bosons).
 
Again one can introduce a "relativistic  atom" comparison hamiltonian.  

It is convenient here to notice that both the given form and the relativistic  atom operator are invariant under rotations and can be decomposed in angular momentum states.

Notice that the expectation value of $ \sqrt { - \Delta}$ depends on the angular momentum of the state. 
 
One can again derive the properties of the spectrum from those of the spectrum of $H_R$.

Denote by $ M_j^b$ the value of $m$ such that for $ m \leq  M_j^b  $ the expectation value of the component of the quadratic form with angular momentum $j$  the form associated to $ \sqrt \Delta  - \frac {C_b (m) }{|x|}$ ceases to be positive.
 
 One verifies that  $1< M _0^b $ whereas $M_j ^b > 1$ for all $ j \geq 1 $   
 
 Again this result is also true for the complete form.
 
 Therefore for two identical bosons of mass 1 which do not interact among themselves and are in contact interaction with a particle of the same mass 1,  in the zero angular momentum sector there are infinitely many extensions.  
 
 In Krein space their negative spectrum is pure  point and not degenerate and is unbounded below with asymptotically a geometric rate. 

There is no interaction in the $l=1$ channel. The quadratic form  is positive for $ l \geq 2 $.

 One must now recall that this analysis is done in Krein space, and to draw conclusions relevant for  physics one must come back to physical space inverting   the Krein map. 

In our case the Krein map  is a change in metric from $L^2 $ to ${ \cal H}^{- \frac{1}{2}}$.

We consider first the case in which the form in Krein space is positive. Since the map preserves positivity, the image in physical space is positive. 

In Krein space the form was strongly closed. By compactness of the Krein map, inverting the map one obtains a form that is weakly closed. 

But  the form is positive and weakly closed and therefore it is also strongly closed. It determines therefore a positive self-adjoint operator in the physical space $ L^2 ( R^6 )$ in the center of mass frame); this  is  the extension we are looking for.

Consider now the case in which there is  in Krein space a family of self-adjoint operators  each of which has a continuous positive spectrum and one or more isolated negative bound states. 

The positive part of the spectrum leads as before  to a positive self-adjoint operator. 

As for the point spectrum, since the family of self-adjoint operators in Krein space \emph{ does not a direct sum decomposition}  the inversion cannot be achieved by acting with the Krein map on each eigenfunction.

One must rely on $ \Gamma$-convergence (after decomposition in sectors of given angular momentum the quadratic form in physical space are locally convex and bounded below).

The (sequential) limit is the lowest  of the family of forms. 

The sequence of negative eigenvalues that diverges geometrically to $ - \infty$ is turned in a sequence  that converge geometrically at zero (Efimov effect).

\bigskip 
\emph{Remark } 

It is worth remarking that an  operator closely related to $H_R $ is introduced  in papers of Theoretical Physics to describe  the three body problem with contact interactions by means of \emph{homogeneous coordinates} and imposing  B-P boundary conditions . 
 
If the masses are such that one has a  multiplicity of extensions (multiplicity of solutions of the Faddeev equations) [P] the singularity of the wave function of the bound states  is apparently found by a self-consistency analysis.  

Notice that the estimate \emph {cannot be obtained by  a perturbative analysis}. 

Passing through Krein space plays in this case the role of the self-consistecy analysis. 

\bigskip
...............

\section{Details for the case of four particles}

The inequality 

\begin{equation}- \sum_{i=1}^4 ( \frac {1}{2} m_i \Delta_i ^2 \geq -\sum_ {i=1}^3 ( \frac {1}{2} m_i \Delta_i) ^2
\end{equation}

(defined as symmetric operator on functions that are supported away from the coincidence hyper-planes) implies that each  interaction term  is mollified more and  the kinetic contribution is larger.  
  
We cannot use the estimates for the three-body case because  there are more (negative) potentials. 

Again in order to study this system we introduce an auxiliary space, this time by means of the operator $ (H_M + \lambda)^{ -  \frac {1}{2}}  $ for the four particle systems.

In this space there are contributions to the quadratic form from the contact terms of each of the three-particle subsystems, slightly modified in Krein space due to the  presence of the fourth particle.

This modification is due to the Krein map (which depends on all four momenta) and disappears  if we undo the Krein map. 
 
Therefore in Krein space the  contribution from  three body subspaces is  analyzed by reducing it to the operator $H_R$. 

\bigskip
We assume here again  that there are enough symmetries so that the kernel of the quadratic form is a function of two variables. 

In the general case the analysis becomes more involved but in any case the kernel of the quadratic form is a sum of kernels each with singularities described  below.

\bigskip
 As already remarked, there is  a  further contribution to the kernel of the quadratic form, due to the interaction between the charge distribution that represent the contact interaction between distinct pairs of particles. 

Proceeding as in the case $N=3$ , one can verify that this term is the sum of a positive form (kinetic  energy) and a form that in position space has kernel which differs for a positive smooth  quadratic form from the quadratic form of the operator 

\begin{equation} 
Z_2 =  \frac {C_2 }{ |y_i - y_j}| 
\end{equation} 

 where $C_2< 0 $ depends on the masses and  $y_k, \; k=1,2 $ are (common) coordinates  of pairs of particles in contact. 

The  singularity in configuration space are the same type  as the one from a three-body subspace  \emph{but in different position in configuration space}  ;  the two contributions can be treated separately.

\bigskip
For a range of masses the sum of the "potential part"  and the "kinetic energy" part  gives a closed  quadratic form; upon inversion of the Krein map the form is only weakly closed but since  it is positive strong closure follows.

For the complementary range the form is no  longer positive and its negative part leads to  a family of self-adjoint extensions with negative point spectrum.

For the positive part of the form the inversion of the Krein map provides again a closed positive form and therefore a self-adjoint operator in physical space.

For the negative part, the inversion of the Krein map  is done by $ \Gamma-$convergence. 

We have already pointed out that \emph{this extension}  is the limit in strong resolvent sense of the hamiltonian $H_0 + V^\epsilon $ when $ \epsilon \to 0$.

\section{Two pair of identical fermions} 

We give now a more detailed analysis of the case of two pairs of identical spin $ \frac {1}{2}$ fermions. 

There are four  resonances but due to  symmetry one has two terms : a term $C_1$ due  the conspiracy  of  two zero-energy resonances  in the subsystem of three particles   and a term $C_2$ due to the interaction of two separate resonances. 

\bigskip
For the case of a pair of identical spin $ \frac {1}{2}$ fermions   of mass one interacting via contact interactions with another pair of spin $ \frac {1}{2}$  of the same mass a computer estimate [MP]  indicates that the resulting quadratic form is positive.  

We   will now show that this is indeed the case and moreover we will prove that the same is true \emph{for an arbitrary number of  identical spin $ \frac {1}{2} $ fermions}. 

Consider  the  system of four spin $ \frac {1}{2} $ fermions   interacting via contact interactions.. 

All particles have the same mass  so that for each three body subsystems the hamiltonian that describes their  contact interaction is  positive.   

The analysis can be easily extended to the case of two pairs of  spin $ \frac {1}{2}$ fermions and also to the case in which the masses of the two fermions are different. 

The generalization to  $N$ spin $ \frac {1}{2} $ fermions will describe the unitary gas. 

\bigskip
 The quadratic form of the full operator  is \emph{in Krein space} the sum a term which represents the kinetic form for four particles  minus the forms associated in Krein space to potentials that have a $ \frac {1}{|y|}$ singularity \emph{but in two different variables}. 

This form in Krein space can more easily be expressed  in Fourier transform (R.Minlos , private communication) but this choice obscures the singularities in configuration space  (as was the case for $N=3$).  

\bigskip
Setting $ \lambda = 0$ the quadratic form is a sum of three terms $ C_0 , C_1, C_2 $  
$$
  (\phi, C_0 \phi)  = 2 \pi^2 \int dk ds \bar \phi (k,s) ,  \sqrt { \frac {3}{4} (k^2 + s^2) +\frac {1}{2} (k,s) } \phi (k,s)
$$
$$
(\phi, C_1 \phi)  = \int dk ds d w \bar \phi (k,w) \frac {\phi (k,w) + \phi (k,s) }{ k^2 + s^2 + w^2 + (k,s) + (k,w) + (s,w) }  
$$
\begin{equation}
(\phi,C_2 \phi)  = - \int dw ds dk \frac {\bar \phi (k,s) \phi (w- \frac {k+s}{2} , -w-\frac{k+s}{2})} { w^2 + \frac {3}{4} (k^2 +s^2)   + \frac {1}{2} (k,s) } 
\end {equation} 

$C_0$ originates  from the lift to Krein space of the kinetic energy,  $C_1 $ originates from the pair of two-body contact interaction between two pairs of particle and  $C_2$ originates from the contact interaction of a triplet of particles  (in presence of the fourth one). 
 
\bigskip
We notice the following properties  

1)

 $C_1$ is zero on functions that are antisymmetric under interchange  of the  two arguments. 
 On these functions $C_0 + C_2$ is the integral with some positive weight of forms of the same type as encountered in the three-body case.
 
 Therefore it is positive if the masses are equal.
 
 \bigskip 
 2)

On functions that are symmetric under interchange $C_2$ is positive and larger that half of the absolute value of $C_1$. 
 
\bigskip  
3)

On the other hand on these  functions $ C_0 + \frac {1}{2} C_1 $ is positive since it is the integral with a positive weight  of kernels that were analyzed in the case $N=3$ and proved to be positive.  

\bigskip
Therefore $ C_0 + C_1 +C_2 $ is the kernel of a positive operator.

This proves that the operator is positive in Krein space. Since the Krein map is positivity preserving we  have  proved

\bigskip 
\emph{Proposition }

\emph{ The operator associated to a system two pairs of identical spin $ \frac {1}{2} $ particles interacting through contact interactions  is a  positive self-adjoint operator in physical space}.

 \bigskip
 ..............

One can verify that the requirement that the two pairs are identical is not needed because the ground state is in any case symmetric.

\bigskip
\emph{Remark } 

We have done our analysis for the non resonant  case. 

But it is easy to see that the resonant case leads to the same results a part from details about the shape of the wave functions and the eigenvalues in the tail of the spectrum in the Efimov effect.) 

The bound states have lower energies. 

\bigskip
...............

\section{ Birman-Schwinger formula for contact interactons}

In quantum mechanics the Birman-Schwinger formula for the difference of two resolvent families  is used to transform a relation between unbounded operators into a relation between two families of bounded operators; it is mostly used to study convergence. 

For regular two-body potentials if we set $ H_\epsilon = H_F + V_\epsilon  $   the Krein map ${\cal K}_\lambda $ is defined by duality  on quadratic forms by  

\begin {equation} 
 {\cal K}_\lambda^\epsilon  (W ) =( \frac { 1}{ H_F + \lambda})^ { \frac {1}{2} } V_\epsilon   ( \frac { 1}{ H_F + \lambda}) ^ { \frac {1}{2} } 
\end{equation}

where $ V_\epsilon $ is the quadratic form of the potential term . 

For smooth potentials the Birman-Schwinger formula for the difference of two resolvent families is

\begin{equation}
\frac {1}{ H_\epsilon  +\lambda } - \frac {1}{H_F +\lambda } = \frac {1}{H_F +\lambda }  W_{\lambda ,\epsilon}      \frac {1}{H_F +\lambda } 
\end{equation} 

where   $  W_{\lambda ,\epsilon} $ is the Birman-Schwinger kernel and $H_\epsilon $ is the total hamiltonian .

For contact interactions elements $ \psi $ in  the domain of $H_\lambda $ can be written as $ \psi = \phi + \xi$ where $ \phi $ is an element the form domain of $H_0$  
 and $ \xi $ square integrable but more singular .

Under this decomposition one has

\begin{equation}
H_\lambda \psi = H^0_\lambda \phi 
\end{equation} 

and the quadratic form of $H$ can be written [A,S] as  

\begin{equation}
 \Phi (\psi_1, \psi_2) = \Phi_0 (\phi_1, \phi_2) ) + \Xi (\xi_1,\xi_2) 
 \end{equation} 
 
 where $ \Phi_0$ is the quadratic form of $H_0$ where  $ \Xi $ is a bilinear form  in the boundary space.

For $ N \geq 3 $ the Birman-Schwinger kernel for the approximating hamiltonian converges in the limit   $ \epsilon \to 0$ (there are no three-body resonances) .  

We can therefore  study  the resolvent  before the limit $ \epsilon \to 0$ \emph{and at this limit } by lifting them   to a  larger Hilbert-Sobolev space (the Krein space) where the "boundary potentials"   have a  finite $L^1$ norm . 

In this space it is easy to verify point-wise convergence of the potentials almost everywhere and then  $L^1$ convergence. 

Also the Birman-Schwinger operators converge. 

In the regular case the convergence is uniform off the real axis.

In the singular case convergence occurs away from the real axis and from the negative spectrum 
of the limit (unique) self-adjoint extension. 

Using the Krein map the Birman-Schwinger formula can be written  

\begin{equation} 
\frac {1}{H + \lambda }  -  \frac {1}{ H_F + \lambda  }  = {\cal K}_\lambda (  {\cal K}_\lambda ( W_\lambda ) ) 
\end{equation} 

where $ W_\lambda = \lim_{\epsilon \to 0 }W_{\lambda, \epsilon}  $ exists in distributional sense due to the smoothing properties of $ {\cal K}_\lambda $.

\bigskip
\emph{Remark}

For contact interactions there is a natural relation between the kernel of the Birman-Schwinger 
kernel and the quadratic form that characterizes the extension.

To see this, notice that the domain is the sum of a part that belongs to the domain of the free laplacian and a more singular part (it correspond in electrostatics  to the contribution of the charges. 

The relation is shown by writing in two different ways the energy form.

Choose $ \lambda $ in such a way that  $ H_\lambda = H+  \lambda I $ is invertible

Introducing the Birman-Schwinger kernel $ W_\lambda $ defined by

\begin{equation} \frac {1}{  H + \lambda} = \frac {1}{ H^0 + \lambda} + \frac {1}{ H^0 + \lambda} 
W_\lambda \frac {1}{ H^0 + \lambda}
\end{equation} 

by (27) one has  

\begin{equation}
(H_\lambda \psi, \frac {1}{H_\lambda } H_\lambda  \psi) =  ( H^0_\lambda \phi , \frac  {1}{ H_\lambda } H^0_\lambda \phi) = (\phi, H^0_\lambda \phi) + (K_\lambda  \phi, K_\lambda    \phi) 
\end{equation}

This relation should be compared with (25)

\section{ Unitary gas,  BEC and BCS}

We have already  noticed the similarity between  the  scaling  of the potential we use with parameter $ \epsilon$ with the scaling $ V_N(x) = N^{3 \beta} V(N^\beta (x))$ which is used in the study of the fluctuations of N-particle quantum systems, for $N$ very large, around the non linear Schr\"odinger equation (see e.g. [B,C,S]) .

Here $ \beta $ is a parameter which can vary between zero and one (from the mean filed regime to the Gross-Pitaewskii regime); the correspondence is $ \epsilon \leftrightarrow \frac {1}{N^{\beta} } $ for $ \beta > 0$.

Our results are for fixed values of $N $ ; this number may be very large. 

In Quantum Mechanics when a parameter takes very large (or small) values, one often has a simpler description when the parameter takes the limit value. 

For interactions of very short range it is natural to use the range $\epsilon $ as a small paramter.

For a system with al large number $N$ of particles it is natural to regard $ \frac {1}{N}$ as a small parameter. 

When considering a gas of very many particles interacting through force of very small range it is instead natural to consider as small parameters both  the range of the interaction and some power of the number of particles.

If only one parameter is used   one can consider several cases according  to the ratio  of the two small paramters. 

This allows to use both techniques related to the Schr\"odinger equation for a many body system with regular potential (of very small but finite range) and Fock space techniques [B,C,S].

On the other hand our choice, i.e to consider first the limit when the support of the potential goes to zero (contact interactions) and then the limit of very large $N$ has some other advantages. 

It permits to prove that the only bound states present for any value of $N$ are three- and four-body bound states (if there are no other interactions).

They may be infinite in number (Efimov effect) but  their binding energy is bounded uniformly.

Therefore the lower bound of spectrum of the hamiltonian of the N-body system with (attractive) contact interactions is at most $ - k N$ for some $k$ that depends on the masses.

This is perhaps the reason why in the analysis of $N$ body systems for $N \to \infty $ and range $ N^{\frac{ \beta}{ \epsilon}} $ with attractive potentials  the potential term is multiplied with an extra factor $\frac {1}{N}$ (see e.g. [B,C,S] ).

The explanation usually given is that this factor is needed to give equal  weight to the kinetic and the potential term.

But in doing so one neglects the landscape of the bound states (which are due to the dominance of the attraction  with respect the kinetic energy). 

\bigskip
Returning to contact interactions  we analyze now the equations for a gas of identical particles with interaction of very short range by taking first the limit of vanishing range and then the limit $ N \to \infty$.

We have stated that the hamiltonian of contact interactions cannot be defined for a system of two \emph{isolated} particles due to an ultraviolet diverge.

Since the diverging term is a \emph{c-number} the Schr\"odinger  equation  for a system of two particles with  point interaction is well defined. 

For $N \geq 3$ both the hamiltonian and the equation is well defined. 

Consider a system of N  particles of mass $m$ which interact through a contact interactions and restrict attention to the configurations  in which two of the particles are so far away from the others that the presence of the other particles  does not effect much their  mutual interaction. 

This is the case for a very diluted gas.

Notice that the equations for contact  interaction (not the hamiltonian) is defined also for two isolated particles. 

If we assume that   the probability of interaction of one of the particles with the other one  is proportional to the probability that the second  particle be there, contact interaction leads in physical space to the non linear equations for the wave function of the two particles 

$$
i \frac {\partial}{\partial t} \phi_1  (t,x)= \frac {1}{m_1} \Delta \phi_1 (t,x) + C|\phi_2  (t,x) | ^2 \phi_1(t,x) \qquad 
$$
\begin{equation}
i \frac {\partial}{\partial t} \phi_2  (t,x)= \frac {1}{m_2}  \Delta \phi_2 (t,x) + C |\phi_1  (t,x) | ^2 \phi _2(t,x) 
\end{equation}

where $C$ is a  coupling constant.

These are the equations of two  particles in contact interaction. 

 Since point interactions are strong resolvent limit of interactions through  short range potentials when the range goes to zero, and we are neglecting the influence of the particles which are far away, \emph{ the solutions of these effective equations} are strong limits, as $\epsilon \to 0$  of  the solutions of the equation for the system of the two selected particles under the compatible assumption  on the initial data.  

Setting  $ |\phi_1 (x) |^2 = |\phi_2 (x) |^2 $ one can regard these equations as effective equations for  a gas of   particles identically distributed if the gas is very  diluted and the interaction is of very small range so small that one can omit configurations in which more than two particles interact. 

The resulting equation \emph{for each particle in the pair} is

\begin{equation}
i \frac {\partial}{\partial t} \phi  (t,x)= \frac {1}{m}  \Delta \phi (t,x) + C |\phi  (t,x) | ^2 \phi (t,x) 
\end{equation}

Notice that one can make use of the compactness of the embedding and standard technology ($\Gamma$- convergence) \emph{also in the   non-linar cases} to transfer the results from the Krein space to the physical space. 

This is particularly relevant if one wants to describe a system of $N$ identical bosons in the limit $ N \to \infty$ through  a non-linear (cubic ) Schr\"odinger equation.

This description holds also for a gas of spin $ \frac {1}{2}$ particles. 
 
 The equation is meant to describe the dynamics of the one-particle marginals for a gas  of identical particles  if the gas is very  diluted and the  range of the interaction is so short that it can be described as a contact  interaction. 
  
The solutions to this equation may therefore be regarded  as describing, in the contact interaction limit,   the solutions of a  \emph{linear equation} for a pair of identically distributed  particles. 

Notice that the resulting equation is non linear equation \emph{for each member of the pair}, but \emph{it is the pair structure that is relevant}. 

If the constant $C$ in front of the non-linear term is the scattering length, passing  from attractive to repulsive forces can be done e.g modulating  a  zero energy resonance (we have seen  that the formalism of point interaction  for $N \geq 3 $ applies also to the case in which there are zero energy resonances). 

The dominant structure is still the interaction between \emph{pairs of particles}  but the properties of the gas change. 

One refers to this transition as transition from B.E.C (Bose Einstein condensation) to B.C.S. (Bardeen, Cooper, Schrieffer) condensate of pairs.

\section {Conclusions}

We have proved that for a Schr\"odinger system of three  or more particles contact interactions are self-adjoint operators which are constructed considering a natural extensions of $H_0$ a symmetric operator defined on functions with support away from coincidence planes s $ \Gamma_{i,j} $ (defined by $ x_i = x_j$).  

These operators are limits, in the strong resolvent sense, of Schr\"odinger hamiltonians with two-body potentials with smaller and smaller support and constant  $L^1 $ norm;  they may have zero-energy resonances. 

The negative (point ) spectrum  of the Hamiltonian for contact interactions is completely determined by the structure of the three-body and four-body subsystems. 

The energies of the bound states  are lower if there are two-body resonances.

Functions  in the  domain of the contact hamiltonian (except bound states) satisfy  B-P boundary conditions at $ \Gamma_{i,j}$. 

An Efimov effect may be present  in the three or four body channels. 

An interesting question that remains open is a more detailed analysis of the role the matrix $B $  of the magnetic terms in originating bound states. 

These terms  are  called  \emph{magnetic term} since in the approximating potential they  depending linearly on the velocity.

\bigskip
\emph{Remark }

Since the only contributions the negative part of the spectrum can be derived from the  three body and four body subsets,  it may be interesting to consider  general N-body structure for contact interaction of spin $ \frac{1}{2}$ fermions. 

We have seen  the four body contribution has two terms; one is a three body term, the other is the sum of a negative three-body term and a positive "two-pairs" contribution. 

The latter  compensates in part the negative one  so that if the matrix $B$ is zero the "kinetic contribution"  makes the system stable.

As  $N $ increases  there are more and more of these terms. 

The sum  is always positive (the system is stable) but the number of  negative and  positive  contributions increases  without bound with $N$. 

Different contributions correspond to conspiracies  between different two-body resonances and they are localized in different regions of configuration space. 

They have different signs and  the                                                                                                                                                                                                                                                                                                                                                                                                                                                                                                                         total contributions to the energy is dominated by the contribution from the kinetic energy. 

If not aware of this compensation ,  one may be led to see the need of a \emph{renormalization} of the negative part.  

\bigskip
Acknowledgments 

Comments and constructive criticism by several friends, in particular A. Michelangeli and K.Yajima,  at an early stage of the research were very helpful and are gratefully acknowledged. I'm also grateful to R.Minlos for valuable correspondence. 

I'm very grateful  to prof. Yajima for the very warm hospitality at Gakushuin University

\bigskip
\centerline{ \emph{References} }

\vskip 4 pt \noindent
[A] S.Albeverio, F.Gesztesy, R.Hoegh-Krohn, H-Holden  Solvable models in Q. M. AMS 2004
\vskip 4 pt \noindent
[A, K] S.Albeverio, P. Kurasov   Lett. in Math. Physics  41 (1997)  79-92
\vskip 4 pt \noindent
[A,F,R] S.albeverio, S.Fassari, F.Rinaldi  J.Phys.A 48 (2015) 185301
\vskip 4 pt \noindent
[A,S] A.Alonso, B.Simon J. Operator Theory 4 (1980) 251-270
\vskip 4 pt \noindent
[B] M.Birman Math. Sb. N.S. 38 (1956) 431-480 
\vskip 4 pt \noindent
[B,P] H. Bethe A.Peierls Proc. Royal Soc. 148 (1935) 146-164
\vskip 4 pt \noindent
[B,C,S] C.Boccato, S.Cenatempo, B Schlein Ann. Henry Poincar 18 (2017) 113-191
\vskip 4 pt \noindent
[Ba,P]  B.Bazak, D.Petrov Phys. Rev Lett. 119  (2017) 083002-083006
\vskip 4 pt \noindent
[B,T] G.Basti, A.Teta  arXiv 1601.08129 (1916) 
\vskip 4 pt \noindent
$[C_1]$ M.Correggi et al  Rev.Math.Phys. 24 (2012) 1250017-32 
\vskip 4 pt \noindent
$[C_2]$ M.Correggiet al. Mathematical Physics, Analysis, Geometry 18(2016) 1--36
\vskip 4 pt \noindent
[C,M,P] Y. Castin, C.Mora, L.Pricoupenko PRL 105 (2016) 2232011-4
\vskip 4 pt \noindent
[C,T]  Y. Castin, E. Tignone Physical Rev A 84 (2011)  062704- 062720
\vskip 4 pt \noindent
[D]  G.F.Dell'Antonio, A.Michelangeli, E .Scandone,  K. Yajima  Ann. Poincar  in press 
\vskip 4 pt \noindent
[Da]  G.S.Danilov Sov. Phys. JETP 13 (1961) 648-660) 
\vskip 4 pt \noindent
[Dal]  G.Dal Maso Introduction to $ \Gamma$-convergence Progr Non Lin. Diff Eq. 8, Birkhauser (1993)
\vskip 4 pt \noindent
[D,R] S.Derezinski, S.Richard ArXiv:1604.03340 
\vskip 4 pt \noindent
[E] V.Efimov Nucler Physics  A 210 (1971) 157-186
\vskip 4 pt \noindent
[E,T] F.Erman, O.Turgut, J.Phys. A (2010) 335204
\vskip 4 pt \noindent
[K] M.G.Krein Rec Math. (Math. Sbornik)   N.S. 20 (1962)1947 431-495
\vskip 4 pt \noindent
[K,R] K.Kowalsy , J.Rembielinski Phys. Rev. A 84 (2011) 012108
\vskip 4 pt \noindent
[lY] A. Le Yaouanc, L. Oliver, J-C Raynar J. of Mathematical Physics 38 (1997) 3998-4012
\vskip 4 pt \noindent
[M] A.Michelangeli The Krei-Visik-Birman theory revisited SISSA preprint 59/2015 
\vskip 4 pt \noindent
[M,F]  R-Minlos, L.Faddeev   Sov. Phys. Doklady 6 (1972) 1072-1074
\vskip 4 pt \noindent
[M,P] A.Michelangeli, P.Pfeiffer   J. PhysicsA  49 (2016) 105301-105331 
\vskip 4 pt \noindent
$ [M_1] $ R.Minlos    Moscow Math. Journal  1 (2011) 111-127
\vskip 4 pt \noindent
$ [M_2]$ R.Minlos  Moscow  Mat. Journal 14 (2014) 617-637
\vskip 4 pt \noindent
[P] A.Posilicano Journ. Functional analysis  183 (2011) 109-147
\vskip 4 pt \noindent
[Pa] B.S.Pavlov  Math. Sbornik 64 (1989) No 1 
\vskip 4 pt \noindent
[Pe]  D.S. Petrov  Physcal Rev Letters 93 (2004) 143201- 143204 
\vskip 4 pt \noindent
[S,B] E.Salpeter, H.Bethe Phys Rev 84 (1952) 1232
\vskip 4 pt \noindent
 [S,T]  G.V.Skorniakov, K. A. Ter-Martirosian Soviet Physics JETP 4 (1957) 648-661
[\vskip 4 pt \noindent
[Y] K. Yajma J.Math. Soc. Japan 47 (1995) 551-581 
\vskip 4 pt \noindent
[W,C] F.Werner, Y.Casten   Phys.Rew. A 74 (2006)  053604- 053650 
\vskip 1 cm \noindent

\bigskip
\section{Appendix : contact interaction vs point interaction}

\emph{Point interaction} and \emph{contact interaction} are two different entities. 

Recall that \emph{point interaction} [A] describes the limit when $ \epsilon \to 0$ of the hamiltonian of a particle of mass $m$  subject to an attractive potential $ V_\epsilon(|x| \equiv \epsilon^{-\frac{3}{2}}V(\frac {|x|}{\epsilon } ) $ under the assumption that $ V \in L^2 (R^3) $ and the system has a zero energy resonance.

We have remarked that, contrary to what occurs for a system of at least three interacting particles, for a two particle system the more natural scaling  $ V_\epsilon (|x|) \equiv  \epsilon^{-3 } V(\frac {|x|}{\epsilon } ) $ (which preserves the $ L^1$ norm) does not produce a limit hamiltonian (but the limit equations exist) . 

The scaling chosen in the definition of point interaction is such that the $L^2$ norm of the potential is independent of $ \epsilon$ and the $L^1$ norm converges to zero when $ \epsilon \to 0$.

The limit is a non-trivial hamiltonian due to the presence of a zero-energy resonance that in the limit  provides a further $ \epsilon^{-\frac{3}{2}}$ factor .

Since the zero energy resonance is a  "large distance behavior "  and the scaling of the potential is a "short distance behavior", the compensation of the two is a non trivial effect; the proof [A] requires a deep 
analysis of the singular behavior of the resolvent at the origin in momentum space. 

In a three body system the  two-body  zero energy resonances play a marginal  role and the "natural scaling" is required in order to obtain a non trivial limit.

It is known that in three dimensions for regular potentials the $ L^p $ to $ L^q$ mapping properties of the Wave Operator are radically different if the there is a zero energy resonance [Y].

For point interaction, also with several fixed points [D] , the mapping properties are the same as in the case of regular potentials with zero-energy resonances. 

In order to find the origin of this similar behavior and also to find a relation between contact and point interactions, it is convenient to remark that  zero energy resonances lead  \emph{a long time scale effect}.

We scale time setting $ \tau = \epsilon^{\frac {3}{2} } t $ so that on the $ \tau $ time scale the displacement  of particle $A$ is  of order $ \epsilon^{-\frac {3}{2} } $.

Since resonances are long distance  effects  it is convenient to consider a particle of very  small mass.

If  moreover one wants  to consider the case in which the positions of some other  particles can be regarded  as  fixed one is led to assume that  these particles are  very  massive. 

Consider an $N+1$  particle systems in which particle $A$ has mass $ \epsilon^{frac{3}{2} }   $ and  the remaining $N $ particles $B_1,.. B_N$  have mass $ \epsilon ^{- \frac {3}{2}} $.

Particle $A$ interacts with the other particles through a potential $V_\epsilon   = \frac {1}{ \epsilon^3}V(\frac {|x|}{\epsilon})$ , $ V(x) \in L^1 (R^3) $ that may admits a zero energy resonance. 

The dynamics is uniquely defined for arbitrary  small values of $ \epsilon$. 

Particles $B_i$ can be considered motionless, i.e located at fixed points $ y_n$.

In this approximation \emph{and on the new time scale}  the Schr\"odinger equation  reads 

\begin{equation} 
 i \frac {\partial}{ \partial \tau } \phi = - \Delta \phi + \epsilon ^{ -\frac{3}{2}} V (\frac {| x-y_i |}{\epsilon})  \phi 
\end{equation} 
 
In the absence of zero energy resonances the dynamics converges in the limit  $ \epsilon \to 0$ to free motion. 

If a zero energy resonance is present,  in the limit $ \epsilon \to 0$ and in the new time scale the dynamics is given  the Sch\"odinger equation  for a particle of unit mass under  \emph{point interaction} with $N$ centers fixed at the point $ y_1, \ldots y_N $.  

Therefore  point interaction with fixed centers can be regarded as an approximation to the asymptotic (in time) dynamics of a particle of very small mass interacting  with particles of large mass through a potential of very short range that gives rise to a zero energy resonance. 

Notice that after a time of order $ \frac {1}{\epsilon^{\frac{3}{2} }} $ the motion is  almost free (ballistic) if the two-body potentials have  no zero energy resonance and is described by a point interactions if there is a zero energy resonance.  

This justifies  the mapping properties of the Wave Operator for point interaction [D].

\end{document}